\documentstyle[12pt]{article}
\newcounter{eq}
\newcounter{sc}

\def\overleftrightarrow#1{\vbox{\ialign{##\crcr
 $\leftrightarrow$\crcr\noalign{\kern-1pt\nointerlineskip}
 $\hfil\displaystyle{#1}\hfil$\crcr}}}

\newlength{\minitwocolumn}
\begin{document}

%%%%%%%%%%%%%%%%%%%%%%%%%%%%%%%%%%%%%%%%%%%%%%%%%%%%%%%%%%%%%%%%%%
%%%%%%%%%%%%%%%%%%%%%%%% Title %%%%%%%%%%%%%%%%%%%%%%%%%%%%%%%%%%%
%%%%%%%%%%%%%%%%%%%%%%%%%%%%%%%%%%%%%%%%%%%%%%%%%%%%%%%%%%%%%%%%%%
\begin{flushright}
DPUR/TH/64\\
September, 2019\\
\end{flushright}
\vspace{20pt}

%\magnification=\magstep1
\pagestyle{empty}
\baselineskip15pt
%\font\cmssB=cmss17
%\font\cmssS=cmss10

\begin{center}
{\large\bf  Planck Scale from Broken Local Conformal Invariance in Weyl Geometry
\vskip 1mm }

\vspace{20mm}

Ichiro Oda\footnote{
           E-mail address:\ ioda@sci.u-ryukyu.ac.jp
                  }

\vspace{10mm}
           Department of Physics, Faculty of Science, University of the 
           Ryukyus,\\
           Nishihara, Okinawa 903-0213, Japan\\

\end{center}

%\maketitle

\vspace{10mm}
\begin{abstract}

It is shown that in the quadratic gravity based on Weyl's conformal geometry, the Planck mass scale can be 
generated from quantum effects of the gravitational field and the Weyl gauge field via the Coleman-Weinberg 
mechanism where a local scale symmetry, that is, conformal symmetry, is broken. At the same time, the Weyl
gauge field acquires mass less than the Planck mass by absorbing the dilaton. The shape of the effective 
potential is almost flat owing to a gravitational character and high symmetries, so our model would provide an
attractive model for the inflationary universe. We also present a toy model showing spontaneous symmetry 
breakdown of a scale symmetry by moving from the Jordan frame to the Einstein one, and point out
its problem.

\end{abstract}

\newpage
\pagestyle{plain}
\pagenumbering{arabic}
%\setcounter{page}{1}

%%%%%%%%%%%%%%%%%%%%%%%%%%%%%%%%%%%%%%%%%%%%%%%%%%%%%%%%%%%%%%%%%%
%%%%%%%%%%%%%%%%%%%%%%%% Article %%%%%%%%%%%%%%%%%%%%%%%%%%%%%%%%%
%%%%%%%%%%%%%%%%%%%%%%%%%%%%%%%%%%%%%%%%%%%%%%%%%%%%%%%%%%%%%%%%%%

\section{Introduction}

One of the most important problems in modern particle physics and cosmology is surely to understand the origin 
of not only the mass of elementary particles but also of different mass scales existing in nature. This understanding 
is also closely related to an understanding of the other important problems such as the gauge hierarchy problem
and the cosmological constant problem. 

In order to understand the origin of the mass and various mass scales, it is natural to start with a theory involving 
no intrinsic mass scales and consider how the mass scales are generated from a massless world via some mechanism.
In this respect, let us recall that the mass, or equivalently the energy, couples to a gravitational field through
the energy-momentum tensor in a universal manner, so one must deal with a gravity directly for understanding the origin
of the mass.  

Another important observation for understanding the origin of the mass is that there naturally appears a local or 
global scale symmetry in theories without intrinsic mass scales.  However, since as stressed in \cite{Oda1}, 
global symmetries are in general against the spirit not only of general relativity owing to no-hair theorem of 
black holes \cite{MTW} but also of superstring theory where gobal symmetries are either explicitly broken or gauged, 
one should work with a gravitational theory which is invariant under not a global 
but a local scale transformation as well as the general coordinate transformation at very short distances.\footnote{In this article, 
we refer to a local scale transformation as conformal transformation.}  

It is well-known, however, that the introduction of a scalar field always makes it possible to construct a conformally
invariant theory from a non-conformally invariant one. For instance, in general relativity, which is never invariant under
conformal transformation, inserting the conformally invariant composite metric, $g_{\mu\nu} \phi(x)^2$, to the Einstein-Hilbert
action gives rise to the conformally invariant scalar-tensor gravity \cite{Fujii}.  

This fact suggests that a conformal symmetry might in fact make no sense while a scale symmetry might play a role in nature. 
We therefore reach an impasse of which symmetry between a global scale symmetry and a local conformal one we should adopt 
as the guiding principle for understanding the origin of the mass: Respecting general relativity and superstring theory forces us to
adopt the conformal symmetry as the guiding principle, but doing the non-triviality leads to paying our attention to the scale 
symmetry. A natural resolution for overcoming this impasse is to couple gauge fields to a theory. Thus, we should consider
a conformally invariant gravitational theory such that conformal symmetry is implemented by a gauge field.     

A conformally invariant extension of general relativity involving a gauge field has been already proposed by Weyl 
on the basis of Weyl's conformal geometry, what we call, the Weyl geometry \cite{Weyl}.\footnote{See Ref. \cite{Scholz} 
for historical review on the Weyl geometry. References related to the Weyl geomerty can be found in Refs. 
\cite{Dirac}-\cite{Ghilencea3}, for instance.}  The Weyl geometry is defined as a geometry equipped with a real symmetric 
metric tensor $g_{\mu\nu}$ as in general relativity and a symmetric connection $\tilde \Gamma^\lambda _{\mu\nu}$, 
which is related to the affine connection $\Gamma^\lambda _{\mu\nu}$ via the relation Eq. (\ref{W-connection}) 
as seen shortly. It turns out that the Weyl geometry reduces to the Riemann geometry when the Weyl gauge field $S_\mu$ 
is vanishing, or more precisely speaking, $S_\mu$ is a gradient, i.e., pure gauge.

In geometrical terms, the Weyl geometry critically differs from the Riemann one in that only angles, but not lengths,
are preserved under parallel transport. To put differently, parallel displacement of a vector field changes its length
in such a way that the very notion of lengths becomes path-dependent. For instance, one can envisage a space traveller, 
who travels to a distant star and then returns to the earth, being surprised to know not only that people in the earth 
have aged much rather than him as predicted by GR in the Riemann geometry but also that the clock on the rocket 
runs at a different rate from that in the earth as understood by Weyl conformal gravity in the Weyl geometry, 
what is called, "the second clock problem" \cite{Penrose}.  Based on this very striking geometry, Weyl has attempted to 
geometrize the electromagnetic theory in the space-time geometry, but failed. This is because the electromagnetic theory 
is described in terms of the $U(1)$ gauge group whereas the Weyl symmetry is generated by a non-compact Abelian
gauge group.

Since we have already applied the idea of generating the Planck scale from radiative corrections via the Coleman-Weinberg 
mechanism \cite{Coleman} to conformally invariant gravitational theories with the SM in the Riemann geometry \cite{Oda1, Oda2}, 
in the article we will turn our attention to a conformally invariant quadratic gravity in the Weyl geometry 
and show that this is indeed the case even in this more general geometry.\footnote{A preliminary work was reported 
in the conference proceeding \cite{Oda3}.}  Concretely, we wish to construct a gravitational theory that satisfies 
the following properties:

\begin{description}
\item[(1)] The classical action is invariant under conformal transformation,
and there are in consequence no fundamental dimensional constants in the classical action.

\item[(2)] The Planck mass scale is dynamically generated from quantum effects through the Coleman-Weinberg
mechanism. 
 
\item[(3)] In the limit of long distances, the effective action approaches the Einstein-Hilbert action with a
cosmological constant for general relativity.
\end{description}

In Section 2, we present a toy model which shows spontaneous symmetry breakdown (SSB) of a scale symmetry.
The key idea is that we begin with a globally scale-invariant action in the Jordan frame and then move to the
Einstein frame. In the process of moving from the Jordan frame to the Einstein frame, we need to introduce
a constant with mass dimension to compensate for the mass dimension of a scalar field, thereby triggering
the SSB of the scale symmetry.  But we also point out a problem of this SSB. In Section 3, we briefly review 
Weyl's conformal geometry. In Section 4, we present a classical action of quadratic gravity in the Weyl geometry.
In Section 5, we calculate the one-loop effective potential in the Coleman-Weinberg formalism. Section 6 is devoted to 
the conclusion. Two appendices account for the spin projection operators in a flat Minkowski space-time, and
the calculation of the functional Jacobian.

\section{Spontaneous symmetry breakdown of scale symmetry} 

There is a well-known mechanism of spontaneous symmetry breakdown of a global scale symmetry,
which can be also extended to the case of a conformal symmetry. In this section, we shall present
a toy model with a scale symmetry, explain why the scale symmetry is broken spontaneously, and then point out 
an unsatisfactory point of this SSB mechanism.

As a toy model, let us work with a scale-invariant Lagrangian density in the Jordan frame, which is defined as
%**   J-Toy model   %%%%%%%%%%%%%%%%%%%%%%%%%%%%%%%%%%%%%%%%%%%%%%%%%%%%%%%%%
\begin{eqnarray}
{\cal{L}} = \sqrt{-g} \left( \frac{1}{2} \xi \phi^2 R - \frac{1}{2} \varepsilon g^{\mu\nu} \partial_\mu \phi
\partial_\nu \phi - \frac{\lambda_1}{4 !} \phi^4 -  \frac{1}{2} g^{\mu\nu} \partial_\mu \Phi
\partial_\nu \Phi - \frac{\lambda_2}{4} \phi^2 \Phi^2 - \frac{\lambda_3}{4 !} \Phi^4 \right),
\label{J-Toy model}
\end{eqnarray}
%%%%%%%%%%%%%%%%%%%%%%%%%%%%%%%%%%%%%%%%%%%%%%%%%%%%%%%%%%%%%%%%%%% 
where $\xi$ is a constant, and $\varepsilon$ takes the value $+1$ for $\phi$ being normal field while it does $-1$
for $\phi$ being ghost field.\footnote{The conformally invariant scalar-tensor gravity is given by $\xi = \frac{1}{6}$ and
$\varepsilon = - 1$.}  Moreover, $\phi$ and $\Phi$ are two distinct scalar fields, and $\lambda_i  (i = 1, 2, 3)$
are dimensionless coupling constants. In this article, we make use of the conventions and notation for the Riemann tensors 
and the metric signature in the Wald textbook \cite{Wald}, and in particular our sign convention is 
$\eta_{\mu\nu} = diag ( - 1, 1, 1, 1)$.

The key step for SSB is to move from the Jordan frame (J-frame) to the Einstein frame (E-frame) by
applying a conformal transformation:
%**   Conformal transf   %%%%%%%%%%%%%%%%%%%%%%%%%%%%%%%%%%%%%%%%%%%%%%%%%%%%%%%%%
\begin{eqnarray}
g_{\mu\nu} \rightarrow g_{\star \mu\nu}  = \Omega (x)^2 g_{\mu\nu}, \qquad
\phi \rightarrow \phi_\star = \Omega (x)^{-1} \phi, \qquad
\Phi \rightarrow \Phi_\star = \Omega (x)^{-1} \Phi,
\label{Conformal transf}
\end{eqnarray}
%%%%%%%%%%%%%%%%%%%%%%%%%%%%%%%%%%%%%%%%%%%%%%%%%%%%%%%%%%%%%%%%%%% 
Moving to the E-frame requires us to choose the scale factor $\Omega(x)$ to
%**   E-frame1   %%%%%%%%%%%%%%%%%%%%%%%%%%%%%%%%%%%%%%%%%%%%%%%%%%%%%%%%%
\begin{eqnarray}
\xi \phi^2 = \Omega(x)^2 M_{Pl}^2, 
\label{E-frame1}
\end{eqnarray}
%%%%%%%%%%%%%%%%%%%%%%%%%%%%%%%%%%%%%%%%%%%%%%%%%%%%%%%%%%%%%%%%%%% 
or equivalently,   
%**   E-frame2   %%%%%%%%%%%%%%%%%%%%%%%%%%%%%%%%%%%%%%%%%%%%%%%%%%%%%%%%%
\begin{eqnarray}
\phi_\star = \frac{M_{Pl}}{\sqrt{\xi}},
\label{E-frame2}
\end{eqnarray}
%%%%%%%%%%%%%%%%%%%%%%%%%%%%%%%%%%%%%%%%%%%%%%%%%%%%%%%%%%%%%%%%%%% 
where  $M_{Pl}$ is the (reduced) Planck mass defined as $M_{Pl} = \frac{1}{\sqrt{8 \pi G}} = 2.44 \times 10^{18} GeV$ 
with $G$ being the Newton constant.
Then, in the E-frame, the Lagrangian density (\ref{J-Toy model}) reduces to the form:
%**   E-Toy model   %%%%%%%%%%%%%%%%%%%%%%%%%%%%%%%%%%%%%%%%%%%%%%%%%%%%%%%%%
\begin{eqnarray}
{\cal{L}} = \sqrt{-g_\star} \left( \frac{M_{Pl}^2}{2} R_\star - \frac{1}{2} g_\star^{\mu\nu} 
\partial_\mu \sigma \partial_\nu \sigma - \frac{\lambda_1}{4 ! \xi^2} M_{Pl}^4 
-  \frac{1}{2} g_\star^{\mu\nu} {\cal{D}}_\mu \Phi_\star {\cal{D}}_\nu \Phi_\star
- \frac{\lambda_2}{4 \xi} M_{Pl}^2 \Phi_\star^2 - \frac{\lambda_3}{4 !} \Phi_\star^4 \right).
\label{E-Toy model}
\end{eqnarray}
%%%%%%%%%%%%%%%%%%%%%%%%%%%%%%%%%%%%%%%%%%%%%%%%%%%%%%%%%%%%%%%%%%% 
Here we have defined
%**   Dilaton   %%%%%%%%%%%%%%%%%%%%%%%%%%%%%%%%%%%%%%%%%%%%%%%%%%%%%%%%%
\begin{eqnarray}
\phi = \frac{M_{Pl}}{\sqrt{\xi}} e^{\frac{\zeta \sigma}{M_{Pl}}}, \qquad
\zeta^{-2} = 6 + \frac{\varepsilon}{\xi}, \qquad
{\cal{D}}_\mu \Phi_\star = \left( \partial_\mu + \frac{\zeta}{M_{Pl}} 
\partial_\mu \sigma \right) \Phi_\star,
\label{Dilaton}
\end{eqnarray}
%%%%%%%%%%%%%%%%%%%%%%%%%%%%%%%%%%%%%%%%%%%%%%%%%%%%%%%%%%%%%%%%%%% 
where a scalar field $\sigma$ is called "dilaton". In this way, it is shown that the scalar
field $\Phi$ becomes massive by "eating" a part of the scalar field $\phi$. This phenomenon
is nothing but spontaneous symmetry breakdown of a scale symmetry.  Of course, before and
after SSB, the number of the physical degrees of freedom remains unchanged.  Incidentally,
the dilaton $\sigma$ would acquire a small mass because of conformal anomaly \cite{Fujii}.  

It is worthwhile to summarize this scenario of the SSB and comment on its problem. First, we have
started with a scale-invariant gravitational theory involving two kinds of scalar fields and only
dimensionless coupling constants. In the process of moving from the J-frame to the E-frame,
we had to introduce a dimensional constant, which is the Planck mass in the present context,
to compensate for the mass dimension of the scalar field. This introduction of the Planck mass
has triggered the SSB of a scale symmetry. Let us note that in the conventional scenario of the SSB, 
there is a potential inducing the SSB whereas we have no such potential in the present SSB.
Nevertheless, the very presence of a solution with dimensional constants justifies the claim 
that the present scenario of the SSB is also nothing but spontaneous symmetry breakdown.  
  
A possible problem, however, arises from the lack of the suitable potential in the sense that
we cannot single out a solution realizing the SSB on the stability argument \cite{Fujii}.
Incidentally, though it might be possible that the cosmological argument would pick up an appropriate vacuum
expectation value (VEV) of a scalar field, it is not plausible that the macroscopic physics 
like cosmology could determine a microscopic configuration such as the VEV.
In order to overcome this problem, we try to derive an effective potential showing the SSB
from radiative corrections of gravitational fields since the appearance of the Planck mass scale
should be connected with quantum gravity.

\section{Weyl conformal geometry} 
 
In this section, we briefly review the basic concepts and definitions of the Weyl conformal geometry.\footnote{See
also Refs. \cite{Fulton, Smolin, Cesare} for a concise introduction of the Weyl geometry.} 
In the Weyl geometry, the Weyl transformation, which is the sum of the local conformal transformation for 
a generic field $\Phi (x)$ and the Weyl gauge transformation for a Weyl gauge field $S_\mu(x)$, is defined as
%**   Weyl transf   %%%%%%%%%%%%%%%%%%%%%%%%%%%%%%%%%%%%%%%%%%%%%%%%%%%%%%%%%
\begin{eqnarray}
\Phi (x) \rightarrow \Phi^\prime (x) = e^{w \Lambda(x)} \Phi (x), \qquad
S_\mu (x) \rightarrow S^\prime_\mu (x) = S_\mu (x) - \frac{1}{f} \partial_\mu \Lambda (x),
\label{Weyl transf}
\end{eqnarray}
%%%%%%%%%%%%%%%%%%%%%%%%%%%%%%%%%%%%%%%%%%%%%%%%%%%%%%%%%%%%%%%%%%% 
where $w$ is the Weyl weight, $f$ is the coupling constant for the Abelian gauge group, and 
$\Lambda(x)$ is the local parameter for the conformal transformation. Writing out the conformal transformation
for various fields explicitly,
%**   Weyl transf 2  %%%%%%%%%%%%%%%%%%%%%%%%%%%%%%%%%%%%%%%%%%%%%%%%%%%%%%%%%
\begin{eqnarray}
g_{\mu\nu} (x) &\rightarrow& g_{\mu\nu}^\prime (x) = e^{2 \Lambda(x)} g_{\mu\nu}(x), \qquad
\phi (x) \rightarrow \phi^\prime (x) = e^{- \Lambda(x)} \phi (x),  \nonumber\\
\psi (x) &\rightarrow& \psi^\prime (x) = e^{- \frac{3}{2} \Lambda(x)} \psi (x), \qquad
A_\mu (x) \rightarrow A^\prime_\mu (x) = A_\mu (x),
\label{Weyl transf 2}
\end{eqnarray}
%%%%%%%%%%%%%%%%%%%%%%%%%%%%%%%%%%%%%%%%%%%%%%%%%%%%%%%%%%%%%%%%%%% 
where $g_{\mu\nu} (x)$, $\phi (x)$, $\psi (x)$ and $A_\mu (x)$ are the metric tensor, scalar, spinor,
and conventional gauge fields, respectively. Here it is convenient to define a Weyl covariant derivative $D_\mu$ for
a generic field $\Phi (x)$ with the Weyl weight $w$ as
%**   W-cov-deriv   %%%%%%%%%%%%%%%%%%%%%%%%%%%%%%%%%%%%%%%%%%%%%%%%%%%%%%%%%
\begin{eqnarray}
D_\mu \Phi \equiv \partial_\mu \Phi + w f S_\mu \Phi,
\label{W-cov-deriv}
\end{eqnarray}
%%%%%%%%%%%%%%%%%%%%%%%%%%%%%%%%%%%%%%%%%%%%%%%%%%%%%%%%%%%%%%%%%%% 
which transforms covariantly under the Weyl transformation:
%**   S-cov-transf   %%%%%%%%%%%%%%%%%%%%%%%%%%%%%%%%%%%%%%%%%%%%%%%%%%%%%%%%%
\begin{eqnarray}
D_\mu \Phi \rightarrow (D_\mu \Phi)^\prime = e^{w \Lambda(x)} D_\mu \Phi.
\label{S-cov-transf}
\end{eqnarray}
%%%%%%%%%%%%%%%%%%%%%%%%%%%%%%%%%%%%%%%%%%%%%%%%%%%%%%%%%%%%%%%%%%% 

As mentioned above, the Weyl geometry is defined as a geometry with a real symmetric metric tensor $g_{\mu\nu}
(= g_{\nu\mu})$ and a symmetric connection $\tilde \Gamma^\lambda_{\mu\nu} (= \tilde \Gamma^\lambda_{\nu\mu})$ 
which is defined as
%**   W-connection   %%%%%%%%%%%%%%%%%%%%%%%%%%%%%%%%%%%%%%%%%%%%%%%%%%%%%%%%%
\begin{eqnarray}
\tilde \Gamma^\lambda_{\mu\nu} &=& \frac{1}{2} g^{\lambda\rho} \left( D_\mu g_{\nu\rho} + D_\nu g_{\mu\rho}
- D_\rho g_{\mu\nu} \right)
\nonumber\\
&=& \Gamma^\lambda_{\mu\nu} + f \left( S_\mu \delta^\lambda_\nu + S_\nu \delta^\lambda_\mu 
- S^\lambda g_{\mu\nu} \right),
\label{W-connection}
\end{eqnarray}
%%%%%%%%%%%%%%%%%%%%%%%%%%%%%%%%%%%%%%%%%%%%%%%%%%%%%%%%%%%%%%%%%%% 
where 
%**   Affine connection   %%%%%%%%%%%%%%%%%%%%%%%%%%%%%%%%%%%%%%%%%%%%%%%%%%%%%%%%%
\begin{eqnarray}
\Gamma^\lambda_{\mu\nu} \equiv \frac{1}{2} g^{\lambda\rho} \left( \partial_\mu g_{\nu\rho} 
+ \partial_\nu g_{\mu\rho} - \partial_\rho g_{\mu\nu} \right),
\label{Affine connection}
\end{eqnarray}
%%%%%%%%%%%%%%%%%%%%%%%%%%%%%%%%%%%%%%%%%%%%%%%%%%%%%%%%%%%%%%%%%%% 
is the affine connection in the Riemann geometry. The most important difference between the Riemann geometry
and the Weyl one is that $\nabla_\lambda g_{\mu\nu} = 0$ (the metric condition) in the Riemann geometry, 
while in the Weyl geometry
%**   W-metric cond   %%%%%%%%%%%%%%%%%%%%%%%%%%%%%%%%%%%%%%%%%%%%%%%%%%%%%%%%%
\begin{eqnarray}
\tilde \nabla_\lambda g_{\mu\nu} \equiv \partial_\lambda g_{\mu\nu} - \tilde \Gamma^\rho_{\lambda\mu} 
g_{\rho\nu} - \tilde \Gamma^\rho_{\lambda\nu} g_{\mu\rho}
= - 2 f S_\lambda g_{\mu\nu}. 
\label{W-metric cond}
\end{eqnarray}
%%%%%%%%%%%%%%%%%%%%%%%%%%%%%%%%%%%%%%%%%%%%%%%%%%%%%%%%%%%%%%%%%%% 
Let us recall that the metric condition implies that length and angle are preserved under parallel transport
where Eq. (\ref{W-metric cond}) does that only angle, but not length, is preserved by the Weyl connection.

Now using the Weyl connection $\tilde \Gamma^\lambda_{\mu\nu}$ one can construct a conformally 
invariant curvature tensor:
%**   W-curv-tensor   %%%%%%%%%%%%%%%%%%%%%%%%%%%%%%%%%%%%%%%%%%%%%%%%%%%%%%%%%
\begin{eqnarray}
\tilde R_{\mu\nu\rho} \, ^\sigma &\equiv& \partial_\nu \tilde \Gamma^\sigma_{\mu\rho} 
- \partial_\mu \tilde \Gamma^\sigma_{\nu\rho} + \tilde \Gamma^\alpha_{\mu\rho} \tilde \Gamma^\sigma_{\alpha\nu} 
- \tilde \Gamma^\alpha_{\nu\rho} \tilde \Gamma^\sigma_{\alpha\mu}
\nonumber\\
&=& R_{\mu\nu\rho} \, ^\sigma + f \left( \delta^\sigma_{[\mu} \nabla_{\nu]} S_\rho 
- \delta^\sigma_\rho \nabla_{[\mu} S_{\nu]} - g_{\rho [\mu} \nabla_{\nu]} S^\sigma \right)
\nonumber\\
&+& f^2 \left( S_{[\mu} \delta^\sigma_{\nu]} S_\rho - S_{[\mu} g_{\nu]\rho} S^\sigma
+ \delta^\sigma_{[\mu} g_{\nu]\rho} S_\alpha S^\alpha \right),
\label{W-curv-tensor}
\end{eqnarray}
%%%%%%%%%%%%%%%%%%%%%%%%%%%%%%%%%%%%%%%%%%%%%%%%%%%%%%%%%%%%%%%%%%% 
where $R_{\mu\nu\rho} \, ^\sigma$ is the curvature tensor in the Riemann geometry, and
we have defined the antisymmetrization by the square bracket, e.g., $A_{[\mu} B_{\nu]} \equiv A_\mu B_\nu
- A_\nu B_\mu$. Then, it is straightforward to prove the following identities:
%**   W-curv-identity   %%%%%%%%%%%%%%%%%%%%%%%%%%%%%%%%%%%%%%%%%%%%%%%%%%%%%%%%%
\begin{eqnarray}
\tilde R_{\mu\nu\rho} \, ^\sigma = - \tilde R_{\nu\mu\rho} \, ^\sigma,  \qquad
\tilde R_{[\mu\nu\rho]} \, ^\sigma = 0, \qquad
\tilde \nabla_{[\lambda} \tilde R_{\mu\nu]\rho} \, ^\sigma = 0.
\label{W-curv-identity}
\end{eqnarray}
%%%%%%%%%%%%%%%%%%%%%%%%%%%%%%%%%%%%%%%%%%%%%%%%%%%%%%%%%%%%%%%%%%% 
The curvature tensor $\tilde R_{\mu\nu\rho} \, ^\sigma$ has $26$ independent components,
twenty of which are possessed by $R_{\mu\nu\rho} \, ^\sigma$ and six by the conformally
invariant field strength $H_{\mu\nu} \equiv \partial_\mu S_\nu - \partial_\nu S_\mu$.

From $\tilde R_{\mu\nu\rho} \, ^\sigma$ it is possible to define a conformally invariant 
Ricci tensor:
%**   W-Ricci-tensor   %%%%%%%%%%%%%%%%%%%%%%%%%%%%%%%%%%%%%%%%%%%%%%%%%%%%%%%%%
\begin{eqnarray}
\tilde R_{\mu\nu} &\equiv& \tilde R_{\mu\rho\nu} \, ^\rho
\nonumber\\
&=& R_{\mu\nu} + f \left( - 2 \nabla_\mu S_\nu - H_{\mu\nu} - g_{\mu\nu} \nabla_{\alpha} S^\alpha \right)
+ 2 f^2 \left( S_\mu S_\nu - g_{\mu\nu} S_\alpha S^\alpha \right).
\label{W-Ricci-tensor}
\end{eqnarray}
%%%%%%%%%%%%%%%%%%%%%%%%%%%%%%%%%%%%%%%%%%%%%%%%%%%%%%%%%%%%%%%%%%% 
Let us note that 
%**   W-Ricci-tensor 2  %%%%%%%%%%%%%%%%%%%%%%%%%%%%%%%%%%%%%%%%%%%%%%%%%%%%%%%%%
\begin{eqnarray}
\tilde R_{[\mu\nu]} \equiv \tilde R_{\mu\nu} - \tilde R_{\nu\mu} = - 4 f H_{\mu\nu}.
\label{W-Ricci-tensor 2}
\end{eqnarray}
%%%%%%%%%%%%%%%%%%%%%%%%%%%%%%%%%%%%%%%%%%%%%%%%%%%%%%%%%%%%%%%%%%% 
Similarly, one can define a conformally not invariant but covariant scalar curvature:
%**   W-scalar-curv   %%%%%%%%%%%%%%%%%%%%%%%%%%%%%%%%%%%%%%%%%%%%%%%%%%%%%%%%%
\begin{eqnarray}
\tilde R \equiv g^{\mu\nu} \tilde R_{\mu\nu} 
= R - 6 f \nabla_\mu S^\mu - 6 f^2 S_\mu S^\mu.
\label{W-scalar-curv}
\end{eqnarray}
%%%%%%%%%%%%%%%%%%%%%%%%%%%%%%%%%%%%%%%%%%%%%%%%%%%%%%%%%%%%%%%%%%% 
We find that under the Weyl transformation, $\tilde R \rightarrow \tilde R^\prime = e^{- 2 \Lambda(x)}
\tilde R$ while $\tilde \Gamma^\lambda_{\mu\nu}, \tilde R_{\mu\nu\rho} \, ^\sigma$ and $\tilde R_{\mu\nu}$
are all invariant.

Finally, we wish to write out a generalization of the Gauss-Bonnet topological invariant in the Weyl geometry 
which can be described as
%**   GB   %%%%%%%%%%%%%%%%%%%%%%%%%%%%%%%%%%%%%%%%%%%%%%%%%%%%%%%%%
\begin{eqnarray}
I_{GB} &\equiv& \int d^4 x \sqrt{-g} \, \epsilon^{\mu\nu\rho\sigma} \epsilon_{\alpha\beta\gamma\delta} \, 
\tilde R_{\mu\nu} \, ^{\alpha\beta} \tilde R_{\rho\sigma} \, ^{\gamma\delta}    
\nonumber\\
&=& - 2 \int d^4 x \sqrt{-g} \,  \left( \tilde R_{\mu\nu\rho\sigma} \tilde R^{\rho\sigma\mu\nu} 
- 4 \tilde R_{\mu\nu} \tilde R^{\nu\mu} + \tilde R^2 - 12 f^2 H_{\mu\nu} H^{\mu\nu} \right)
\nonumber\\
&=& - 2 \int d^4 x \sqrt{-g} \,  \left( R_{\mu\nu\rho\sigma} R^{\mu\nu\rho\sigma} 
- 4 R_{\mu\nu} R^{\mu\nu} + R^2 \right).   
\label{GB}
\end{eqnarray}
%%%%%%%%%%%%%%%%%%%%%%%%%%%%%%%%%%%%%%%%%%%%%%%%%%%%%%%%%%%%%%%%%%% 

\section{Quadratic gravity in Weyl geometry} 

In this section, we will present a gravitational theory on the basis of the Weyl geometry outlined 
in the previous section. It is of interest to notice that if only the metric tensor is allowed to use for 
the construction of a gravitational action, the action invariant under the Weyl transformation must be 
of form of quadratic gravity, but not be of the Einstein-Hilbert type. Using the topological invariant (\ref{GB}), 
one can write out a general action of quadratic gravity, which is invariant under the Weyl transformation, as follows:
%**   QG   %%%%%%%%%%%%%%%%%%%%%%%%%%%%%%%%%%%%%%%%%%%%%%%%%%%%%%%%%
\begin{eqnarray}
S_{QG} = \int d^4 x \sqrt{-g} \left[ - \frac{1}{2 \xi^2}  \tilde C_{\mu\nu\rho\sigma} \tilde C^{\mu\nu\rho\sigma} 
+ \frac{\lambda}{4 !} \tilde R^2 \right] \equiv \int d^4 x \sqrt{-g} \, {\cal{L}}_{QG},
\label{QG}
\end{eqnarray}
%%%%%%%%%%%%%%%%%%%%%%%%%%%%%%%%%%%%%%%%%%%%%%%%%%%%%%%%%%%%%%%%%%% 
where $\xi$ and $\lambda$ are dimensionless coupling constants. And a generalization of the conformal tensor, 
$\tilde C_{\mu\nu\rho\sigma}$, in the Weyl geometry is defined as in $C_{\mu\nu\rho\sigma}$ in the Riemann geometry:
%**   Conformal tensor   %%%%%%%%%%%%%%%%%%%%%%%%%%%%%%%%%%%%%%%%%%%%%%%%%%%%%%%%%
\begin{eqnarray}
\tilde C_{\mu\nu\rho\sigma} &\equiv& \tilde R_{\mu\nu\rho\sigma} - \frac{1}{2} \left( g_{\mu\rho} \tilde R_{\nu\sigma}
+ g_{\nu\sigma} \tilde R_{\mu\rho} -  g_{\mu\sigma} \tilde R_{\nu\rho} - g_{\nu\rho} \tilde R_{\mu\sigma} \right)
+ \frac{1}{6} \left( g_{\mu\rho} g_{\nu\sigma} - g_{\mu\sigma} g_{\nu\rho} \right) \tilde R
\nonumber\\
&=& C_{\mu\nu\rho\sigma} + f \left[ - g_{\rho\sigma} H_{\mu\nu} + \frac{1}{2} \left( g_{\mu\rho} H_{\nu\sigma}
+  g_{\nu\sigma} H_{\mu\rho} - g_{\mu\sigma} H_{\nu\rho} - g_{\nu\rho} H_{\mu\sigma} \right) \right].
\label{Conformal tensor}
\end{eqnarray}
%%%%%%%%%%%%%%%%%%%%%%%%%%%%%%%%%%%%%%%%%%%%%%%%%%%%%%%%%%%%%%%%%%%  
This conformal tensor in the Weyl geometry has the following properties:
%**   Conformal tensor 2  %%%%%%%%%%%%%%%%%%%%%%%%%%%%%%%%%%%%%%%%%%%%%%%%%%%%%%%%%
\begin{eqnarray}
\tilde C_{\mu\nu\rho\sigma} = - \tilde C_{\nu\mu\rho\sigma}, \qquad 
\tilde C_{\mu\nu\rho} \, ^\nu = 0, \qquad 
\tilde C_{\mu\nu\rho} \, ^\rho = - 4 f H_{\mu\nu}.
\label{Conformal tensor 2}
\end{eqnarray}
%%%%%%%%%%%%%%%%%%%%%%%%%%%%%%%%%%%%%%%%%%%%%%%%%%%%%%%%%%%%%%%%%%%  

Next, by introducing a scalar field $\phi$ and using the classical equivalence, it is possible to rewrite $\tilde R^2$ 
in the action (\ref{QG}) in the form of the scalar-tensor gravity plus $\lambda \phi^4$ interaction \cite{Ghilencea2} 
whose Lagrangian density takes the form
%**   QG 2  %%%%%%%%%%%%%%%%%%%%%%%%%%%%%%%%%%%%%%%%%%%%%%%%%%%%%%%%%
\begin{eqnarray}
\frac{1}{\sqrt{-g}} {\cal{L}}_{QG} &=& - \frac{1}{2 \xi^2}  \tilde C_{\mu\nu\rho\sigma} \tilde C^{\mu\nu\rho\sigma} 
+ \frac{\lambda}{12} \phi^2 \tilde R - \frac{\lambda}{4 !} \phi^4 
\nonumber\\
&=& - \frac{1}{2 \xi^2}  \tilde C_{\mu\nu\rho\sigma} \tilde C^{\mu\nu\rho\sigma} 
+ \frac{1}{12} \phi^2 \tilde R - \frac{\lambda_\phi}{4 !} \phi^4 
\nonumber\\
&=& - \frac{1}{2 \xi^2}  C_{\mu\nu\rho\sigma} C^{\mu\nu\rho\sigma} 
+ \frac{1}{12} \phi^2 R - \frac{\lambda_\phi}{4 !} \phi^4 - \frac{3 f^2}{\xi^2} H_{\mu\nu}^2 
\nonumber\\
&-& \frac{1}{2} \phi^2 ( f \nabla_\mu S^\mu + f^2 S_\mu S^\mu ),  
\label{QG 2}
\end{eqnarray}
%%%%%%%%%%%%%%%%%%%%%%%%%%%%%%%%%%%%%%%%%%%%%%%%%%%%%%%%%%%%%%%%%%%  
where in the second equality we have redefined $\sqrt{\lambda} \phi \rightarrow \phi$ and set 
$\lambda = \frac{1}{\lambda_\phi}$. 

Here we wish to comment on one remark. The equation motion for $\phi$ in Eq. (\ref{QG 2})
gives us the equation 
%**   phi-eq   %%%%%%%%%%%%%%%%%%%%%%%%%%%%%%%%%%%%%%%%%%%%%%%%%%%%%%%%%
\begin{eqnarray}
\phi^2 = \tilde R = R - 6 f \nabla_\mu S^\mu - 6 f^2 S_\mu S^\mu,
\label{phi-eq}
\end{eqnarray}
%%%%%%%%%%%%%%%%%%%%%%%%%%%%%%%%%%%%%%%%%%%%%%%%%%%%%%%%%%%%%%%%%%% 
where we have used Eq. (\ref{W-scalar-curv}). As shown in the next section, it turns out that
$\langle \phi^2 \rangle \sim M_{Pl}^2$. Thus, at low energies, having a flat Minkowski metric satisfying
$R = 0$ as a classical solution demands that the Weyl gauge field $S_\mu$ takes some suitable classical 
value which breaks the Lorentz invariance or a nonzero condensation $\langle S_\mu S^\mu \rangle \neq 0$ 
in order to cancel $M_{Pl}^2$.  Within the present framework, it seems to be difficult at least technically 
to have a condensation of the Weyl gauge field, $\langle S_\mu S^\mu \rangle \neq 0$, so we are led to assume 
that $S_\mu$ takes a value which breaks the Lorentz invariance. It is not clear at present whether such a
Lorentz-noninvariant solution is admissible as a classical solution or not except the case of cosmology.

Finally, by introducing matter fields, it is straightforward to write down a standard model (SM) or physics
beyond the standard model (BSM) action which is invariant under the Weyl transformation, but we will omit to 
tough on it in this article and present the detail in a separate publication.\footnote{Some related models on the
basis of the Weyl geometry have been made in Refs. \cite{Cheng, Nishino}.}

\section{Emergence of Planck scale}

At low energies, general relativity (GR) describes various gravitational and astrophysical phenomena neatly,
so the Weyl invariant Lagrangian density (\ref{QG 2}) of quadratic gravity should be reduced to that of GR at low
energies. To do that, we need to break the Weyl symmetry at any rate by some method. One method is to 
appeal to the procedure of spontaneous symmetry breakdown (SSB) explained in terms of a toy model in Section 2.
However, as emphasized there, since there is no potential to induce this SSB in the theory, we have no idea which
solution we should pick up among many of configurations from the stability argument. 

The other simple procedure is to take a gauge condition for the Weyl transformation such that $\phi = \phi_0$ where $\phi_0$ 
is a certain constant \cite{Smolin, Cesare, Ghilencea1}. However, $\phi_0$ is a free parameter which is not fixed 
from the stability argument of the potential either so it is not clear why we choose a specific value $\phi_0 \sim M_{Pl}$ 
where $M_{Pl}$ is the Planck mass scale.     

In this article, we would like to look for an alternative possibility by considering a conformally invariant gravitational theory 
where the scalar field $\phi$ acquires a vacuum expectation value (VEV) as a result of instabilities in the full quantum theory 
including quantum corrections from gravity. It is reasonable to conjecture that quantum gravity plays a role in
generating the Planck mass scale dynamically since effects of quantum gravity are more dominant than the other interactions
around the Planck energy. Technically speaking, what we expect is that after quantum corrections of gravitational fields 
are taken into consideration the effective potential has a form favoring the specific VEV, $\langle \phi \rangle \sim M_{Pl}$.
Since effects of quantum gravity make a contribution to the generation of the effective potential, it is natural to
think that the specific VEV takes the value around the Planck mass scale. 

To this aim, let us first expand the scalar field and the metric around a classical field $\phi_c$ and a flat Minkowski
metric $\eta_{\mu\nu}$ like \cite{Oda1}
%**   Expansion   %%%%%%%%%%%%%%%%%%%%%%%%%%%%%%%%%%%%%%%%%%%%%%%%%%%%%%%%%
\begin{eqnarray}
\phi = \phi_c + \varphi,    \quad 
g_{\mu\nu} = \eta_{\mu\nu} + \xi h_{\mu\nu},
\label{Expansion}
\end{eqnarray}
%%%%%%%%%%%%%%%%%%%%%%%%%%%%%%%%%%%%%%%%%%%%%%%%%%%%%%%%%%%%%%%%%%% 
where we take $\phi_c$ to be a constant since we are interested in the effective potential depending on
the constant $\phi_c$. Let us note that unlike the standard assignment of dimensions, $h_{\mu\nu}$ is
now a dimensionless field since $g_{\mu\nu}$ and $\xi$ also have no mass dimension.   
Next, since we wish to calculate the one-loop effective potential, we will
derive only quadratic terms in quantum fields from the classical Lagrangian density (\ref{QG 2}).
Then, the Lagrangian density corresponding to the conformal tensor squared takes the form
%**   Weyl Lagr   %%%%%%%%%%%%%%%%%%%%%%%%%%%%%%%%%%%%%%%%%%%%%%%%%%%%%%%%%
\begin{eqnarray}
{\cal L}_C \equiv - \frac{1}{2 \xi^2}  \sqrt{-g} \, C_{\mu\nu\rho\sigma} C^{\mu\nu\rho\sigma} 
= - \frac{1}{4} h^{\mu\nu} P^{(2)}_{\mu\nu, \rho\sigma} \Box^2 h^{\rho\sigma},
\label{Weyl Lagr}
\end{eqnarray}
%%%%%%%%%%%%%%%%%%%%%%%%%%%%%%%%%%%%%%%%%%%%%%%%%%%%%%%%%%%%%%%%%%%
where $P^{(2)}_{\mu\nu, \rho\sigma}$ is the projection operator for spin-2 modes\footnote{We follow
the definition of projection operators in \cite{Nakasone1, Nakasone2}. The detail is explained in the appendix A.} 
and $\Box \equiv \eta^{\mu\nu} \partial_\mu \partial_\nu$. In a similar manner, the Lagrangian density 
corresponding to the scalar-tensor gravity in Eq. (\ref{QG 2}) reads
%**   ST-Lagr   %%%%%%%%%%%%%%%%%%%%%%%%%%%%%%%%%%%%%%%%%%%%%%%%%%%%%%%%%
\begin{eqnarray}
{\cal L}_{ST} &\equiv& \sqrt{-g} \, \frac{1}{12} \phi^2 R
\nonumber\\
&=& \frac{1}{48} \xi^2 \phi^2_c h^{\mu\nu} \left( P^{(2)}_{\mu\nu, \rho\sigma} 
- 2 P^{(0, s)}_{\mu\nu, \rho\sigma} \right) \Box h^{\rho\sigma} 
- \frac{1}{6} \xi \phi_c \varphi \left( \eta_{\mu\nu} - \frac{1}{\Box} \partial_\mu \partial_\nu \right) 
\Box h^{\mu\nu}.
\label{ST-Lagr}
\end{eqnarray}
%%%%%%%%%%%%%%%%%%%%%%%%%%%%%%%%%%%%%%%%%%%%%%%%%%%%%%%%%%%%%%%%%%% 
The remaining Lagrangian density can be evaluated in a similar way and consequently all the quadratic terms 
in (\ref{QG 2}) are summarized to
%**   Total-Lagr   %%%%%%%%%%%%%%%%%%%%%%%%%%%%%%%%%%%%%%%%%%%%%%%%%%%%%%%%%
\begin{eqnarray}
{\cal L}^{(2)} &=& \frac{1}{4} h^{\mu\nu} \left[ \left( - \Box + \frac{1}{12} \xi^2 \phi^2_c \right)
P^{(2)}_{\mu\nu, \rho\sigma} - \frac{1}{6} \xi^2 \phi^2_c P^{(0, s)}_{\mu\nu, \rho\sigma} \right]
\Box h^{\rho\sigma} 
\nonumber\\
&-& \frac{1}{6} \xi \phi_c \varphi \left( \eta_{\mu\nu} 
- \frac{1}{\Box} \partial_\mu \partial_\nu \right) \Box h^{\mu\nu}
- \frac{\lambda_\phi}{4} \phi_c^2 \varphi^2 -  \frac{\lambda_\phi}{12} \xi \phi_c^3 h \varphi
+ \frac{1}{96} \lambda_\phi \xi^2 \phi_c^4 h_{\mu\nu}^2 
\nonumber\\
&-& \frac{1}{192} \lambda_\phi \xi^2 \phi_c^4 h^2 - \frac{1}{4} H^{\prime \ 2}_{\mu\nu} 
- \frac{1}{24} \xi^2 \phi_c^2 S^\prime_\mu S^{\prime \mu} - \frac{1}{2} \varphi \Box \varphi,
\label{Total-Lagr}
\end{eqnarray}
%%%%%%%%%%%%%%%%%%%%%%%%%%%%%%%%%%%%%%%%%%%%%%%%%%%%%%%%%%%%%%%%%%% 
where we have defined $h= \eta^{\mu\nu} h_{\mu\nu}$ and set $S^\prime_\mu = \frac{2 \sqrt{3} f}{\xi}
 ( S_\mu - \frac{1}{f \phi_c} \partial_\mu \varphi)$  and $H^\prime_{\mu\nu} = \partial_\mu S^\prime_\nu 
- \partial_\nu S^\prime_\mu$. 

At this stage, it is convenient to use the York decomposition for the metric fluctuation field $h_{\mu\nu}$
\cite{York}:
%**   York   %%%%%%%%%%%%%%%%%%%%%%%%%%%%%%%%%%%%%%%%%%%%%%%%%%%%%%%%%
\begin{eqnarray}
h_{\mu\nu} &=& h_{\mu\nu}^{TT} + \partial_\mu \xi_\nu + \partial_\nu \xi_\mu + \partial_\mu \partial_\nu \sigma
- \frac{1}{4} \eta_{\mu\nu} \Box \sigma + \frac{1}{4} \eta_{\mu\nu} h \nonumber\\
&=& h_{\mu\nu}^{TT} + \partial_\mu \xi_\nu + \partial_\nu \xi_\mu
+ \frac{1}{4} \theta_{\mu\nu} s + \frac{1}{4} \omega_{\mu\nu} w,
\label{York}
\end{eqnarray}
%%%%%%%%%%%%%%%%%%%%%%%%%%%%%%%%%%%%%%%%%%%%%%%%%%%%%%%%%%%%%%%%%%%
where $h_{\mu\nu}^{TT}$ is both transverse and traceless, and $\xi_\mu$ is transverse:
%**   TT   %%%%%%%%%%%%%%%%%%%%%%%%%%%%%%%%%%%%%%%%%%%%%%%%%%%%%%%%%
\begin{eqnarray}
\partial^\mu h_{\mu\nu}^{TT} = \eta^{\mu\nu} h_{\mu\nu}^{TT} = \partial^\mu \xi_\mu = 0.
\label{TT}
\end{eqnarray}
%%%%%%%%%%%%%%%%%%%%%%%%%%%%%%%%%%%%%%%%%%%%%%%%%%%%%%%%%%%%%%%%%%% 
Moreover, we have defined 
%**   s&w   %%%%%%%%%%%%%%%%%%%%%%%%%%%%%%%%%%%%%%%%%%%%%%%%%%%%%%%%%
\begin{eqnarray}
s = h - \Box \sigma, \quad  w = h + 3 \Box \sigma, \quad 
\theta_{\mu\nu} =  \eta_{\mu\nu} - \frac{1}{\Box} \partial_\mu \partial_\nu, \quad
\omega_{\mu\nu} = \frac{1}{\Box} \partial_\mu \partial_\nu.
\label{s&w}
\end{eqnarray}
%%%%%%%%%%%%%%%%%%%%%%%%%%%%%%%%%%%%%%%%%%%%%%%%%%%%%%%%%%%%%%%%%%% 
Using the York decomposition (\ref{York}), the Lagrangian density (\ref{Total-Lagr}) reads
%**   Total-Lagr2   %%%%%%%%%%%%%%%%%%%%%%%%%%%%%%%%%%%%%%%%%%%%%%%%%%%%%%%%%
\begin{eqnarray}
{\cal L}^{(2)} &=& \frac{1}{4} h^{TT \mu\nu} \left[ ( - \Box + m^2 ) \Box 
+ \frac{m^2}{2} \lambda_\phi \phi^2_c \right] h^{TT}_{\mu\nu}
- \frac{m^2}{4} \lambda_\phi \phi^2_c \hat \xi_\mu \hat \xi^\mu 
- \frac{1}{4} H^{\prime \ 2}_{\mu\nu} - \frac{m^2}{2} S^\prime_\mu S^{\prime \mu}
\nonumber\\
&-& \frac{1}{2} \varphi ( \Box + \frac{1}{2} \lambda_\phi \phi^2_c ) \varphi
- \frac{3 m^2}{32} h ( \Box + \frac{1}{3} \lambda_\phi \phi^2_c ) h
- \frac{3 m^2}{32} \hat \sigma ( \Box - \lambda_\phi \phi^2_c ) \hat \sigma
+ \frac{3 m^2}{16} h \Box \hat \sigma 
\nonumber\\
&-& \frac{\sqrt{3} m}{4} \varphi \Box ( h - \hat \sigma )
- \frac{\sqrt{3} m}{6} \lambda_\phi \phi^2_c h \varphi, 
\label{Total-Lagr2}
\end{eqnarray}
%%%%%%%%%%%%%%%%%%%%%%%%%%%%%%%%%%%%%%%%%%%%%%%%%%%%%%%%%%%%%%%%%%% 
where we have put $m^2 = \frac{1}{12} \xi^2 \phi_c^2$, and then introduced 
the dimensionless fields $\hat \xi_\mu = \sqrt{\Box} \xi_\mu$ and $\hat \sigma = \Box \sigma$
like $h_{\mu\nu}$. 
 
In order to diagonalize the terms involving the fields $\varphi, h$ and $\hat \sigma$, it is convenient to
assume that
%**   Assumption   %%%%%%%%%%%%%%%%%%%%%%%%%%%%%%%%%%%%%%%%%%%%%%%%%%%%%%%%%
\begin{eqnarray}
\lambda_\phi \propto \xi^4 \ll 1,
\label{Assumption}
\end{eqnarray}
%%%%%%%%%%%%%%%%%%%%%%%%%%%%%%%%%%%%%%%%%%%%%%%%%%%%%%%%%%%%%%%%%%% 
and work with the perturbation series in $\lambda_\phi$. We will prove later that our assumption (\ref{Assumption}) 
is self-consistent and there are no large logarithms. With the assumption (\ref{Assumption}), let us change 
variables from $(\varphi, h, \hat \sigma)$ to $(\varphi^\prime, h^\prime, \hat \sigma^\prime)$
%**   Change-variable   %%%%%%%%%%%%%%%%%%%%%%%%%%%%%%%%%%%%%%%%%%%%%%%%%%%%%%%%%
\begin{eqnarray}
\varphi &=& \varphi^\prime - \frac{\sqrt{3} m}{4} \left[ \left( 1 + \frac{1}{6} \lambda_\phi \phi^2_c 
\frac{1}{\Box} \right) h^\prime - \frac{2}{3} \left( 1 - \frac{5}{6} \lambda_\phi \phi^2_c 
\frac{1}{\Box} \right) \hat \sigma^\prime \right],
\nonumber\\
h &=& h^\prime + \frac{1}{3} \hat \sigma^\prime,  \qquad
\hat \sigma = \hat \sigma^\prime. 
\label{Change-variable}
\end{eqnarray}
%%%%%%%%%%%%%%%%%%%%%%%%%%%%%%%%%%%%%%%%%%%%%%%%%%%%%%%%%%%%%%%%%%% 
It turns out that the associated Jacobian factor is $1$. Then, taking the leading-order terms in $\lambda_\phi$
in each term, the quadratic Lagrangian density (\ref{Total-Lagr2}) can be rewritten to the form
%**   Total-Lagr3   %%%%%%%%%%%%%%%%%%%%%%%%%%%%%%%%%%%%%%%%%%%%%%%%%%%%%%%%%
\begin{eqnarray}
{\cal L}^{(2)} &=& \frac{1}{4} h^{TT \mu\nu} ( - \Box + m^2 ) \Box h^{TT}_{\mu\nu}
- \frac{m^2}{4} \lambda_\phi \phi^2_c \hat \xi_\mu \hat \xi^\mu 
- \frac{1}{4} H^{\prime \ 2}_{\mu\nu} - \frac{m^2}{2} S^\prime_\mu S^{\prime \mu}
\nonumber\\
&-& \frac{1}{2} \varphi^\prime \Box \varphi^\prime
+ \frac{5 m^2}{192} \lambda_\phi \phi^2_c (h^\prime)^2
- \frac{5 m^2}{72} \lambda_\phi \phi^2_c (\hat \sigma^\prime)^2.
\label{Total-Lagr3}
\end{eqnarray}
%%%%%%%%%%%%%%%%%%%%%%%%%%%%%%%%%%%%%%%%%%%%%%%%%%%%%%%%%%%%%%%%%%% 

Next let us set up the gauge-fixing conditions. For the general coordinate transformation and the Weyl transformation, 
we adopt gauge conditions, respectively
%**   Gauge-diffo   %%%%%%%%%%%%%%%%%%%%%%%%%%%%%%%%%%%%%%%%%%%%%%%%%%%%%%%%%
\begin{eqnarray}
\partial^\nu h_{\mu\nu} = \sqrt{\Box} \hat \xi_\mu + \frac{1}{4} \partial_\mu ( h^\prime
+ \frac{10}{3} \hat \sigma^\prime ) = 0, \qquad
\partial_\mu S^{\prime \mu} = 0.
\label{Gauge-diffo}
\end{eqnarray}
%%%%%%%%%%%%%%%%%%%%%%%%%%%%%%%%%%%%%%%%%%%%%%%%%%%%%%%%%%%%%%%%%%%
The corresponding FP ghost terms are respectively calculated to
%**   FP   %%%%%%%%%%%%%%%%%%%%%%%%%%%%%%%%%%%%%%%%%%%%%%%%%%%%%%%%%
\begin{eqnarray}
\det \Delta_{FP}^{(GCT)} = \det ( \Box \delta_\mu^\nu + \partial_\mu \partial^\nu ), \qquad
\det \Delta_{FP}^{(Weyl)} = \det ( \Box ).
\label{FP}
\end{eqnarray}
%%%%%%%%%%%%%%%%%%%%%%%%%%%%%%%%%%%%%%%%%%%%%%%%%%%%%%%%%%%%%%%%%%%
Then, the partition function of the present theory is given by
%**   Partition   %%%%%%%%%%%%%%%%%%%%%%%%%%%%%%%%%%%%%%%%%%%%%%%%%%%%%%%%%
\begin{eqnarray}
&{}& Z [ \phi_c ] = \int {\cal{D}} g_{\mu\nu} {\cal{D}} \phi {\cal{D}} S_\mu \det \Delta_{FP}^{(GCT)}
\det \Delta_{FP}^{(Weyl)} \delta(\partial^\nu h_{\mu\nu}) \delta(\partial_\mu S^{\prime \mu}) 
\exp i \int d^4 x {\cal L}^{(2)}
\nonumber\\
&=& \int {\cal{D}} h^{TT}_{\mu\nu} {\cal{D}} \hat \xi_\mu {\cal{D}} \hat \sigma^\prime 
{\cal{D}} h^\prime {\cal{D}} \varphi^\prime {\cal{D}} S^\prime_\mu \det ( \Box \delta_\mu^\nu 
+ \partial_\mu \partial^\nu ) \det (\Box) 
\delta \left( \sqrt{\Box} \hat \xi_\mu + \frac{1}{4} \partial_\mu ( h^\prime
+ \frac{10}{3} \hat \sigma^\prime ) \right) 
\nonumber\\
&\times& \delta(\partial_\mu S^{\prime \mu})
\exp \ i \int d^4 x \Biggl[ \frac{1}{4} h^{TT \mu\nu} ( - \Box + m^2 ) \Box h^{TT}_{\mu\nu}
- \frac{m^2}{4} \lambda_\phi \phi^2_c \hat \xi_\mu \hat \xi^\mu 
- \frac{1}{2} \varphi^\prime \Box \varphi^\prime
\nonumber\\
&+& \frac{5 m^2}{192} \lambda_\phi \phi^2_c (h^\prime)^2
- \frac{5 m^2}{72} \lambda_\phi \phi^2_c (\hat \sigma^\prime)^2
- \frac{1}{2} S_\mu^\prime ( - \Box + m^2 ) S^{\prime \mu} 
+ \frac{1}{2} \left( \partial_\mu S^{\prime \mu} \right)^2 \Biggr]
\nonumber\\
&=& \frac{\det( \Box \delta_\mu^\nu 
+ \partial_\mu \partial^\nu ) \det (\Box)}{(\det_\xi \Box)^{\frac{1}{2}} (\det_{\varphi^\prime} \Box)^{\frac{1}{2}}
(\det_{h^{TT}} ( - \Box + m^2 ) \Box )^{\frac{1}{2}} (\det_{S^\prime} ( - \Box + m^2 )^{\frac{1}{2}}}.
\label{Partition}
\end{eqnarray}
%%%%%%%%%%%%%%%%%%%%%%%%%%%%%%%%%%%%%%%%%%%%%%%%%%%%%%%%%%%%%%%%%%% 

Using the partition function (\ref{Partition}), we can evaluate the one-loop effective action by integrating out 
quantum fluctuations.  Then, up to a classical potential, recalling the definition $m^2 = \frac{1}{12} \xi^2 \phi^2_c$
the effective action $\Gamma [\phi_c]$ reads
%**   EA   %%%%%%%%%%%%%%%%%%%%%%%%%%%%%%%%%%%%%%%%%%%%%%%%%%%%%%%%%
\begin{eqnarray}
\Gamma [\phi_c] = - i \log Z[\phi_c] = i \frac{5 + 3}{2} \log \mathrm{det} \left( - \Box + \frac{1}{12} \xi^2 \phi^2_c \right).
\label{EA}
\end{eqnarray}
%%%%%%%%%%%%%%%%%%%%%%%%%%%%%%%%%%%%%%%%%%%%%%%%%%%%%%%%%%%%%%%%%%% 
Here two remarks are in order. First, in this expression, the factors $5$ and $3$ come from the fact that 
a massive spin-2 graviton and a massive spin-1 Weyl gauge field possess five and three physical degrees of freedom, respectively. 
Second, let us note that we have ignored the part of the effective action which is independent of $\phi_c$ since it never
gives us the effective potential for $\phi_c$.

To calculate $\Gamma [\phi_c]$, we will proceed step by step: First, let us note that $\Gamma [\phi_c]$ can be 
rewritten as follows:
%**   EA2   %%%%%%%%%%%%%%%%%%%%%%%%%%%%%%%%%%%%%%%%%%%%%%%%%%%%%%%%%
\begin{eqnarray}
\Gamma [\phi_c] &=& 4 i \, \mathrm{Tr} \log \left( - \Box + \frac{1}{12} \xi^2 \phi^2_c \right)
\nonumber\\
&=& 4 i \int d^4 x \, \langle x|  \log \left( - \Box + \frac{1}{12} \xi^2 \phi^2_c \right) | x \rangle
\nonumber\\
&=& 4 i \int d^4 x \int \frac{d^4 k}{(2 \pi)^4} \, \langle x|  \log \left( - \Box + \frac{1}{12} \xi^2 \phi^2_c \right) 
| k \rangle \langle k | x \rangle
\nonumber\\
&=& 4 i (VT) \int \frac{d^4 k}{(2 \pi)^4} \log \left( k^2 + \frac{1}{12} \xi^2 \phi^2_c \right)
\nonumber\\
&=& 4 (VT) \frac{\Gamma(- \frac{d}{2})}{(4 \pi)^{\frac{d}{2}}} \left( \frac{1}{12} \xi^2 \phi^2_c \right)^{\frac{d}{2}},
\label{EA2}
\end{eqnarray}
%%%%%%%%%%%%%%%%%%%%%%%%%%%%%%%%%%%%%%%%%%%%%%%%%%%%%%%%%%%%%%%%%%% 
where $(VT)$ denotes the space-time volume and in the last equality we have used the Wick rotation and 
the dimensional regularization.

Next, let us evaluate $\Gamma [\phi_c]$ in terms of the modified minimal subtraction scheme. In this scheme, 
the $\frac{1}{\varepsilon}$ poles (where $\varepsilon \equiv 4 - d$) together with the Euler-Mascheroni constant
$\gamma$ and $\log (4 \pi)$ are subtracted and then replaced with $\log M^2$ where $M$ is an arbitrary
mass parameter which is introduced to make the final equation dimensionally correct \cite{Peskin}. 
By subtracting the $\frac{1}{\varepsilon}$ pole, (\ref{EA2}) is reduced to the form
%**   EA3   %%%%%%%%%%%%%%%%%%%%%%%%%%%%%%%%%%%%%%%%%%%%%%%%%%%%%%%%%
\begin{eqnarray}
- \frac{1}{VT} \Gamma [\phi_c] &=& - 4 \frac{\Gamma( 2 - \frac{d}{2} )}{\frac{d}{2} (\frac{d}{2} - 1)}
\frac{1}{(4 \pi)^{\frac{d}{2}}} \left( \frac{1}{12} \xi^2 \phi^2_c \right)^{\frac{d}{2}}
\nonumber\\
&=& - \frac{4}{2 (4 \pi)^2} \left( \frac{1}{12} \xi^2 \phi^2_c \right)^2 \left[ \frac{2}{\varepsilon} - \gamma
+ \log (4 \pi)  - \log \left(\frac{1}{12} \xi^2 \phi^2_c \right) + \frac{3}{2} \right]
\nonumber\\
&\rightarrow& \frac{2}{(4 \pi)^2} \left( \frac{1}{12} \xi^2 \phi^2_c \right)^2 
\left[ \log \left(\frac{\xi^2 \phi^2_c}{12 M^2} \right) - \frac{3}{2} \right].
\label{EA3}
\end{eqnarray}
%%%%%%%%%%%%%%%%%%%%%%%%%%%%%%%%%%%%%%%%%%%%%%%%%%%%%%%%%%%%%%%%%%% 
Then, the one-loop effective potential will be of form\footnote{At first sight, the existence of the
$c_2 \phi^2$ term might appear to be strange, but this term in fact emerges in the cutoff regularization. Note that
the only logarithmically divergent term, but not quadratic divergent one, arises in the dimensional regularization.}
%**   EP   %%%%%%%%%%%%%%%%%%%%%%%%%%%%%%%%%%%%%%%%%%%%%%%%%%%%%%%%%
\begin{eqnarray}
V_{eff}^{(1)} (\phi_c) = c_1 + c_2 \phi^2 + \frac{1}{1152 \pi^2} \xi^4 \phi^4_c \log \left(\frac{\phi^2_c}{c_3}\right),
\label{EP}
\end{eqnarray}
%%%%%%%%%%%%%%%%%%%%%%%%%%%%%%%%%%%%%%%%%%%%%%%%%%%%%%%%%%%%%%%%%%% 
where $c_i (i = 1, 2, 3)$ are constants to be determined by the renormalization conditions:
%**   Ren-cond   %%%%%%%%%%%%%%%%%%%%%%%%%%%%%%%%%%%%%%%%%%%%%%%%%%%%%%%%%
\begin{eqnarray}
\left. V_{eff}^{(1)} \right\vert_{\phi_c = 0} = \left. \frac{d^2 V_{eff}^{(1)}}{d \phi^2_c} \right\vert_{\phi_c = 0} 
= \left. \frac{d^4 V_{eff}^{(1)}}{d \phi^4_c} \right\vert_{\phi_c = \mu} = 0,
\label{Ren-cond}
\end{eqnarray}
%%%%%%%%%%%%%%%%%%%%%%%%%%%%%%%%%%%%%%%%%%%%%%%%%%%%%%%%%%%%%%%%%%%
where $\mu$ is the renormalization mass. As a result, we have the one-loop effective potential
%**   EP2   %%%%%%%%%%%%%%%%%%%%%%%%%%%%%%%%%%%%%%%%%%%%%%%%%%%%%%%%%
\begin{eqnarray}
V_{eff}^{(1)} (\phi_c) = \frac{1}{1152 \pi^2} \xi^4 \phi^4_c \left( \log \frac{\phi^2_c}{\mu^2} - \frac{25}{6} \right).
\label{EP2}
\end{eqnarray}
%%%%%%%%%%%%%%%%%%%%%%%%%%%%%%%%%%%%%%%%%%%%%%%%%%%%%%%%%%%%%%%%%%% 
Finally, by adding the classical potential we can arrive at the effective potential in the one-loop approximation
%**   EP3   %%%%%%%%%%%%%%%%%%%%%%%%%%%%%%%%%%%%%%%%%%%%%%%%%%%%%%%%%
\begin{eqnarray}
V_{eff} (\phi_c) = \frac{\lambda_\phi}{4 !} \phi^4_c + \frac{1}{1152 \pi^2} \xi^4 \phi^4_c 
\left( \log \frac{\phi^2_c}{\mu^2} - \frac{25}{6} \right).
\label{EP3}
\end{eqnarray}
%%%%%%%%%%%%%%%%%%%%%%%%%%%%%%%%%%%%%%%%%%%%%%%%%%%%%%%%%%%%%%%%%%% 

It is easy to see that this effective potential has a minimum at $\phi_c = \langle \phi \rangle$ away from the origin 
where the effective potential, $V_{eff} (\langle \phi \rangle)$, is negative. Since the renormalization mass $\mu$ 
is arbitrary, we will choose it to be the actual location of the minimum, $\mu = \langle \phi \rangle$ \cite{Coleman}:
%**   EP4   %%%%%%%%%%%%%%%%%%%%%%%%%%%%%%%%%%%%%%%%%%%%%%%%%%%%%%%%%
\begin{eqnarray}
V_{eff} (\phi_c) = \frac{\lambda_\phi}{4 !} \phi^4_c + \frac{1}{1152 \pi^2} \xi^4 \phi^4_c 
\left( \log \frac{\phi^2_c}{\langle \phi \rangle^2} - \frac{25}{6} \right).
\label{EP4}
\end{eqnarray}
%%%%%%%%%%%%%%%%%%%%%%%%%%%%%%%%%%%%%%%%%%%%%%%%%%%%%%%%%%%%%%%%%%% 
Since $\phi_c = \langle \phi \rangle$ is defined to be the minimum of $V_{eff}$, we deduce
%**   Min-cond   %%%%%%%%%%%%%%%%%%%%%%%%%%%%%%%%%%%%%%%%%%%%%%%%%%%%%%%%%
\begin{eqnarray}
0 &=& \left. \frac{d V_{eff}}{d \phi_c} \right\vert_{\phi_c = \langle \phi \rangle}
\nonumber\\ 
&=& \left( \frac{\lambda_\phi}{6} - \frac{11}{864 \pi^2} \xi^4 \right) \langle \phi \rangle^3,
\label{Min-cond}
\end{eqnarray}
%%%%%%%%%%%%%%%%%%%%%%%%%%%%%%%%%%%%%%%%%%%%%%%%%%%%%%%%%%%%%%%%%%% 
or equivalently, 
%**   Min-cond2   %%%%%%%%%%%%%%%%%%%%%%%%%%%%%%%%%%%%%%%%%%%%%%%%%%%%%%%%%
\begin{eqnarray}
\lambda_\phi = \frac{11}{144 \pi^2} \xi^4.
\label{Min-cond2}
\end{eqnarray}
%%%%%%%%%%%%%%%%%%%%%%%%%%%%%%%%%%%%%%%%%%%%%%%%%%%%%%%%%%%%%%%%%%% 
This relation is similar to $\lambda = \frac{33}{8 \pi^2} e^4$ in case of the scalar QED in 
Ref. \cite{Coleman}, so as in that paper, the perturbation theory holds for very small $\xi$.
Furthermore, it is worthwhile to point out that Eq.  (\ref{Min-cond2}) guarantees our previous 
assumption (\ref{Assumption}). 

The substitution of Eq. (\ref{Min-cond2}) into $V_{eff}$ in (\ref{EP4}) yields
%**   EP5   %%%%%%%%%%%%%%%%%%%%%%%%%%%%%%%%%%%%%%%%%%%%%%%%%%%%%%%%%
\begin{eqnarray}
V_{eff} (\phi_c) = \frac{1}{1152 \pi^2} \xi^4 \phi^4_c 
\left( \log \frac{\phi^2_c}{\langle \phi \rangle^2} - \frac{1}{2} \right).
\label{EP5}
\end{eqnarray}
%%%%%%%%%%%%%%%%%%%%%%%%%%%%%%%%%%%%%%%%%%%%%%%%%%%%%%%%%%%%%%%%%%% 
Thus, the effective potential is now parametrized in terms of $\xi$ and $\langle \phi \rangle$
instead of $\xi$ and $\lambda_\phi$; it is nothing but the well-known "dimensional transmutation", 
i.e., a dimensionless coupling constant $\lambda_\phi$ is traded for a dimensional quantity $\langle \phi \rangle$ 
via symmetry breakdown of the $\it{local}$ Weyl symmetry.  

A peculiar feature of the effective potential (\ref{EP5}) is that the overall coefficient $\frac{1}{1152 \pi^2}$ is 
much smaller than that of the scalar QED, $\frac{3}{64 \pi^2}$ \cite{Coleman} owing to a gravitational character 
and high symmetries. This feature means that compared to the scalar QED, the shape of our effective potential (\ref{EP5}) 
is almost flat so that it would provide an attractive model for the inflationary cosmology. Actually, we could
propose the following interesting cosmological scenario: The 'dilaton' $\varphi$ rolls down toward the potential minimum 
$\langle \phi \rangle$ along the potential (\ref{EP5}) very slowly from the high energy region more than the Planck energy, 
during which the universe inflates, and then it is trapped at the minimum $\langle \phi \rangle$ around which
the dilaton oscillates very violently generating matter fields of the SM and afterwards it is absorbed into
the Weyl gauge field, by which the Weyl gauge field becomes massive. A remakable point of this scenario is that
we can kill two birds with one stone: The massive Weyl gauge field is a strong candidate for dark matter since
it is known that the Weyl gauge field does not interact with matter fields but has only the gravitational 
interaction \cite{Shirafuji, Hayashi}. In addition to it, the dilaton, which is a dangerous source of the fifth force,
completely disappears from the particle spectrum by being eaten by the Weyl gauge field.          

Hence, from the classical Lagrangian density (\ref{QG 2}) of quadratic gravity, via dimensional transmutation,
the Einstein-Hilbert term for GR is induced in such a way that the Planck mass $M_{Pl}$ is given by
%**   Planck mass   %%%%%%%%%%%%%%%%%%%%%%%%%%%%%%%%%%%%%%%%%%%%%%%%%%%%%%%%%
\begin{eqnarray}
M_{Pl}^2 = \frac{1}{6} \langle \phi \rangle^2.
\label{Planck mass}
\end{eqnarray}
%%%%%%%%%%%%%%%%%%%%%%%%%%%%%%%%%%%%%%%%%%%%%%%%%%%%%%%%%%%%%%%%%%% 
At the same time, the Weyl gauge field becomes massive by 'eating' the dilaton $\varphi$ whose magnitude 
of mass is given 
%**   Gauge mass   %%%%%%%%%%%%%%%%%%%%%%%%%%%%%%%%%%%%%%%%%%%%%%%%%%%%%%%%%
\begin{eqnarray}
m_S^2 = \frac{1}{12} \xi^2 \langle \phi \rangle^2 = \frac{1}{2} \xi^2 M_{Pl}^2.
\label{Gauge mass}
\end{eqnarray}
%%%%%%%%%%%%%%%%%%%%%%%%%%%%%%%%%%%%%%%%%%%%%%%%%%%%%%%%%%%%%%%%%%% 
As regards the dilaton, properly speaking, $\log \varphi$ is in fact proportional to the dilaton field as mentioned in Section 2.
Moreover, it is not manifest in the present formulation that the 'dilaton' $\varphi$ is eaten by the Weyl gauge
field since there remains the kinetic term of $\varphi^\prime$ in (\ref{Total-Lagr3}). This is because of
our choice of the gauge-fixing condition, $\partial_\mu S^{\prime \mu} = 0$. To see the mass spectrum more clearly, 
it is necessary to move from the Lorenz gauge to the "unitary gauge", i.e., $\phi = \sqrt{6} M_{Pl}$, for which we have 
no dilaton in the mass spectrum.

Finally, as long as the perturbation theory is concerned, the coupling constant $\xi$ must take a small value,
$\xi \ll 1$, so that the mass size of the massive Weyl gauge field is less than the Planck mass. For instance, if we assume
that $\xi^2 \sim 10^{-4}$ as in the QED, $\alpha = \frac{e^2}{4 \pi} \sim 10^{-4}$, the mass size of the Weyl gauge mass 
is around the GUT scale, and the Weyl gauge field in essence decouples at low enegies, thereby making it possible to 
avoid the second clock problem \cite{Ghilencea1, Ghilencea2, Ghilencea3}. In other words, at the low energy region 
satisfying $E \ll m_S$, we can integrate over the massive Weyl gauge field, and consequently not only we would have GR 
(with the SM) but also the second clock effect does not has physical effects any longer.

\section{Conclusions}

Shortly after Einstein has established general relativity (GR) in 1915, in order to unify gravity and electromagnetic
interaction Weyl has advocated a generalization of GR in that the very notion of length becomes path-dependent. 
In Weyl's theory, even if the lightcones also retain the fundamental role as in GR, there is no absolute meaning 
of scales for space-time, so the metric is defined only up to proportionality. It is this property that we have 
a scale symmetry prohibiting the appearance of any dimensionful parameters and coupling constants 
in the Weyl theory. The main complaint against the Weyl's idea is that it inevitably leads to the so-called "second clock 
effect" \cite{Penrose}: The rate where any clock measures would depend on its history. Since the second clock effect 
has not been observed by experiments, the Weyl theory might make no sense as a classical theory. 

However, viewed as a quantum field theory, the Weyl theory is a physically consistent theory and provides us with
a natural playground for constructing conformally invariant quantum field theories as shown in this article.\footnote{We
have already contructed the other scale invariant gravitational models \cite{Oda4, Oda5, Oda6}.}
Requiring the invariance under Weyl transformation is so strong that only quadratic curvature terms are allowed to
exist in a classical action, which should be contrasted with the situation of GR where any number of curvature terms could be 
in principle the candidate of a classical action only if we require the action to be invariant under diffeomorphisms.

We have a lot of problems to be clarified in future.  One serious problem is the problem of unitarity. The lack of perturbative
unitarity is a common problem in the higher derivative gravity theories \cite{Stelle}. However, it is expected that 
like the conformal gravity \cite{Tonin, Fradkin} whose Lagrangian density is of form, $\sqrt{-g} C_{\mu\nu\rho\sigma}^2$, 
the Weyl's quadratic gravity at hand might be asymptotically free so that the issue of the perturbative unitarity is closely 
relevant to infrared dynamics of asymptotic fields, and as a result this problem would become quite nontrivial. 
Provided that we can confine the ghosts in the Weyl gravity like in QCD, we would be free of the perturbative unitarity. 
The other interesting problems will be conformal anomaly, the gauge hierarchy problem and the cosmological constant. 
We wish to return to these problems in future.

%%%%%%%%%%%%%%%%%%%%%%%%%%%%%%%%%%%%%%%%%%%%%%%%%%%%%%%%%%%%%%%%%%
%%%%%%%%%%%%%%%%%%%%%%%% Acknowledgements %%%%%%%%%%%%%%%%%%%%%%%%%%%%%
%%%%%%%%%%%%%%%%%%%%%%%%%%%%%%%%%%%%%%%%%%%%%%%%%%%%%%%%%%%%%%%%%%
\begin{flushleft}
{\bf Acknowledgements}
\end{flushleft}

We would like to thank T. Kugo for valuable discussions.

%%%%%%%%%%%%%%%%%%%%%%%%%%%%%%%%%%%%%%%%%%%%%%%%%%%%%%%%%%%%%%%%%%
%%%%%%%%%%%%%%%%%%%%%%%% Appendix %%%%%%%%%%%%%%%%%%%%%%%%%%%%%
%%%%%%%%%%%%%%%%%%%%%%%%%%%%%%%%%%%%%%%%%%%%%%%%%%%%%%%%%%%%%%%%%%
\appendix  

\section{Projection operators}

In a flat Minkowski space-time, it is often convenient to split a field into irreducible representation of 
the Lorentz group where each component corresponds to a degree of freedom of different spin.
For instance, a symmetric rank-2 tensor field can be decomposed into four irreducible components
of the Lorentz group, which correspond to spin-2, spin-1 and two spin-0 degrees of freedom. 

It is then customary to introduce the following projection operators in the generic $d$-dimensions
\cite{Nakasone1, Nakasone2}:
%**   Spin-proj1   %%%%%%%%%%%%%%%%%%%%%%%%%%%%%%%%%%%%%%%%%%%%%%%%%%%%%%%%%
\begin{eqnarray}
P^{(2)}_{\mu\nu, \rho\sigma} &=& \frac{1}{2} ( \theta_{\mu\rho} \theta_{\nu\sigma} 
+ \theta_{\mu\sigma} \theta_{\nu\rho} ) -  \frac{1}{d-1} \theta_{\mu\nu} \theta_{\rho\sigma},
\nonumber\\
P^{(1)}_{\mu\nu, \rho\sigma} &=& \frac{1}{2} ( \theta_{\mu\rho} \omega_{\nu\sigma} 
+ \theta_{\mu\sigma} \omega_{\nu\rho} + \theta_{\nu\rho} \omega_{\mu\sigma} 
+ \theta_{\nu\sigma} \omega_{\mu\rho} ),
\nonumber\\
P^{(0, s)}_{\mu\nu, \rho\sigma} &=& \frac{1}{d-1} \theta_{\mu\nu} \theta_{\rho\sigma},
\qquad
P^{(0, w)}_{\mu\nu, \rho\sigma} = \omega_{\mu\nu} \omega_{\rho\sigma},
\nonumber\\
P^{(0, sw)}_{\mu\nu, \rho\sigma} &=& \frac{1}{\sqrt{d-1}} \theta_{\mu\nu} \omega_{\rho\sigma},
\qquad
P^{(0, ws)}_{\mu\nu, \rho\sigma} = \frac{1}{\sqrt{d-1}} \omega_{\mu\nu} \theta_{\rho\sigma},
\label{Spin-proj1}
\end{eqnarray}
%%%%%%%%%%%%%%%%%%%%%%%%%%%%%%%%%%%%%%%%%%%%%%%%%%%%%%%%%%%%%%%%%%% 
where the transverse operator $\theta_{\mu\nu}$ and the longitudinal operator $\omega_{\mu\nu}$
are defined as
%**   T-L operators   %%%%%%%%%%%%%%%%%%%%%%%%%%%%%%%%%%%%%%%%%%%%%%%%%%%%%%%%%
\begin{eqnarray}
\theta_{\mu\nu} = \eta_{\mu\nu} - \frac{1}{\Box} \partial_\mu \partial_\nu 
= \eta_{\mu\nu} - \omega_{\mu\nu},
\qquad
\omega_{\mu\nu} = \frac{1}{\Box} \partial_\mu \partial_\nu.
\label{T-L operators}
\end{eqnarray}
%%%%%%%%%%%%%%%%%%%%%%%%%%%%%%%%%%%%%%%%%%%%%%%%%%%%%%%%%%%%%%%%%%% 
These spin projection operators satisfy the orthogonality relations
%**   Ortho-rel   %%%%%%%%%%%%%%%%%%%%%%%%%%%%%%%%%%%%%%%%%%%%%%%%%%%%%%%%%
\begin{eqnarray}
P^{(i, a)}_{\mu\nu, \rho\sigma} P^{(j, b)}_{\rho\sigma, \lambda\tau} &=& \delta^{ij} \delta^{ab}  
P^{(i, a)}_{\mu\nu, \lambda\tau},
\qquad
P^{(i, ab)}_{\mu\nu, \rho\sigma} P^{(j, cd)}_{\rho\sigma, \lambda\tau} = \delta^{ij} \delta^{bc}  
P^{(i, a)}_{\mu\nu, \lambda\tau},
\nonumber\\
P^{(i, a)}_{\mu\nu, \rho\sigma} P^{(j, bc)}_{\rho\sigma, \lambda\tau} &=& \delta^{ij} \delta^{ab}  
P^{(i, ac)}_{\mu\nu, \lambda\tau},
\qquad
P^{(i, ab)}_{\mu\nu, \rho\sigma} P^{(j, c)}_{\rho\sigma, \lambda\tau} = \delta^{ij} \delta^{bc}  
P^{(i, ac)}_{\mu\nu, \lambda\tau},
\label{Ortho-rel}
\end{eqnarray}
%%%%%%%%%%%%%%%%%%%%%%%%%%%%%%%%%%%%%%%%%%%%%%%%%%%%%%%%%%%%%%%%%%% 
with $i, j = 0, 1, 2$ and $a, b, c, d = s, w$ and the completeness relation
%**   Com-rel   %%%%%%%%%%%%%%%%%%%%%%%%%%%%%%%%%%%%%%%%%%%%%%%%%%%%%%%%%
\begin{eqnarray}
[ P^{(2)} + P^{(1)} + P^{(0, s)} + P^{(0, w)} ]_{\mu\nu, \rho\sigma} 
= \frac{1}{2} ( \eta_{\mu\rho} \eta_{\nu\sigma} + \eta_{\mu\sigma} \eta_{\nu\rho} ).
\label{Com-rel}
\end{eqnarray}
%%%%%%%%%%%%%%%%%%%%%%%%%%%%%%%%%%%%%%%%%%%%%%%%%%%%%%%%%%%%%%%%%%% 
Together with the York decomposition (\ref{York}), the projection operators give rise to the relations:
%**   Spin-proj2   %%%%%%%%%%%%%%%%%%%%%%%%%%%%%%%%%%%%%%%%%%%%%%%%%%%%%%%%%
\begin{eqnarray}
P_{\mu\nu}^{(2) \rho\sigma} h_{\rho\sigma} &=& h_{\mu\nu}^{TT}, \qquad 
P_{\mu\nu}^{(1) \rho\sigma} h_{\rho\sigma} = \partial_\mu \xi_\nu + \partial_\nu \xi_\mu,
\nonumber\\
P_{\mu\nu}^{(0, s) \rho\sigma} h_{\rho\sigma} &=&  \frac{1}{4} \theta_{\mu\nu} s, \qquad
P_{\mu\nu}^{(0, w) \rho\sigma} h_{\rho\sigma} =  \frac{1}{4} \omega_{\mu\nu} w.
\label{Spin-proj2}
\end{eqnarray}
%%%%%%%%%%%%%%%%%%%%%%%%%%%%%%%%%%%%%%%%%%%%%%%%%%%%%%%%%%%%%%%%%%% 
 
\section{Functional Jacobian}
In this appendix, we wish to present the calculation of the functional Jacobian associated with the change 
of variables, $h_{\mu\nu} \rightarrow (h^{TT}_{\mu\nu}, \hat \xi_\mu, \hat \sigma, h)$. To do that, 
we will use the relation \cite{Percacci2}
%**   Measure   %%%%%%%%%%%%%%%%%%%%%%%%%%%%%%%%%%%%%%%%%%%%%%%%%%%%%%%%%
\begin{eqnarray}
1 = \int {\cal{D}} h_{\mu\nu} e^{ - {\cal{G}}(h, h) },
\label{Measure}
\end{eqnarray}
%%%%%%%%%%%%%%%%%%%%%%%%%%%%%%%%%%%%%%%%%%%%%%%%%%%%%%%%%%%%%%%%%%%
where ${\cal{G}}(h, h)$ is an inner product in the space of symmetric rank-2 tensors:
%**   G   %%%%%%%%%%%%%%%%%%%%%%%%%%%%%%%%%%%%%%%%%%%%%%%%%%%%%%%%%
\begin{eqnarray}
{\cal{G}}(h, h) &=& \int d^4 x \left( h_{\mu\nu} h^{\mu\nu} + \frac{a}{2} h^2 \right)  \nonumber\\
&=& \int d^4 x \left[ (h^{TT}_{\mu\nu})^2 - 2 \hat \xi_\mu \hat \xi^\mu 
+ \left( \frac{1}{4} + \frac{a}{2} \right) h^2  + \frac{3}{4} \hat \sigma^2 \right],
\label{G}
\end{eqnarray}
%%%%%%%%%%%%%%%%%%%%%%%%%%%%%%%%%%%%%%%%%%%%%%%%%%%%%%%%%%%%%%%%%%%
where $a$ is an arbitrary constant except $a = - \frac{1}{2}$. Thus, the functional Jacobian $J$ 
which is defined as
%**   J   %%%%%%%%%%%%%%%%%%%%%%%%%%%%%%%%%%%%%%%%%%%%%%%%%%%%%%%%%
\begin{eqnarray}
{\cal{D}} h_{\mu\nu} = J  {\cal{D}} h^{TT}_{\mu\nu} {\cal{D}} \hat \xi_\mu {\cal{D}} \hat \sigma {\cal{D}} h,
\label{J}
\end{eqnarray}
%%%%%%%%%%%%%%%%%%%%%%%%%%%%%%%%%%%%%%%%%%%%%%%%%%%%%%%%%%%%%%%%%%%
is given by
%**   J2   %%%%%%%%%%%%%%%%%%%%%%%%%%%%%%%%%%%%%%%%%%%%%%%%%%%%%%%%%
\begin{eqnarray}
J  = 1.
\label{J2}
\end{eqnarray}
%%%%%%%%%%%%%%%%%%%%%%%%%%%%%%%%%%%%%%%%%%%%%%%%%%%%%%%%%%%%%%%%%%%

%%%%%%%%%%%%%%%%%%%%%%% reference %%%%%%%%%%%%%%%%%%%%%%%%%%%%%%%
%%%%%%%%%%%%%%%%%%%%%%%%%%%%%%%%%%%%%%%%%%%%%%%%%%%%%%%%%%%%%%%%%%

\end{document}